\documentclass[11pt,a4paper]{article}

\usepackage{amssymb}
\usepackage{amsmath}

\newcommand{\be}{\begin{equation}}
\newcommand{\ee}{\end{equation}}
\newcommand{\bea}{\begin{eqnarray}}
\newcommand{\eea}{\end{eqnarray}}
\newcommand{\lb}{\label}
\newcommand{\bdm}{\begin{displaymath}}
\newcommand{\edm}{\end{displaymath}}

\begin{document}


\noindent
\begin{center}
\vspace*{1cm}

{\large\bf DOES TIME EXIST IN QUANTUM GRAVITY?} 

\vskip 1cm

{\bf Claus Kiefer} 
\vskip 0.4cm
Institute for Theoretical Physics,\\ University of Cologne, \\
Z\"ulpicher Strasse~77,
50937 K\"oln, Germany.\\ {\tt http://www.thp.uni-koeln.de/gravitation/}
\vspace{1cm}

\begin{abstract}

Time is absolute in standard quantum theory and dynamical in general
relativity. The combination of both theories into a theory of quantum
gravity leads therefore to a ``problem of time''. In my essay I shall
investigate those consequences for the concept of
time that may be drawn without a detailed knowledge of quantum gravity. The only
assumptions are the experimentally supported universality of the linear
structure of quantum theory and the recovery of general relativity in
the classical limit.
Among the consequences are the fundamental
timelessness of quantum gravity, the approximate nature of a
semiclassical time, and the correlation of entropy with
the size of the Universe.
\end{abstract}

\end{center}



\section{Time in Physics}

On December 14, 1922, Albert Einstein delivered a speech to students
and faculty members of Kyoto University in which he summarized how he
created his theories of relativity \cite{Einstein}. As for the key
idea in finding special relativity in 1905, he emphasized: ``An analysis of
the concept of time was my solution.'' He was 
then able to complete his theory within five weeks. 

An analysis of the concept of time may also be the key for the
construction of a quantum theory of gravity. Such a hope is supported
by the fact that a change of the fundamental equations in physics is
often accompanied by a change in the notion of time. Let me briefly
review the history of time in physics.  

Before Newton, and thus before the advent of modern science, time was
associated with periodic motion, notably the motion of the
`Heavens'. It was therefore a countable time, each tick corresponding
to one period; there was no idea of a continuum. 

It was Newton's great achievement to invent the notion of an absolute
and continuous time. Such a concept was needed for the formulation of
his laws of mechanics and universal gravitation. Although Newton's
concepts of absolute space and absolute time were heavily criticized
by some contemporaries as being unobservable, alternative relational
formulations were only constructed after the advent of general
relativity in the 20th century \cite{Barbour}.

In Einstein's theory of special relativity, time was unified with
space to form a four-dimensional spacetime. But this ``Minkowski
spacetime'' still constitutes an absolute background in the sense that
there is no {\em reactio} of fields and matter -- Minkowski spacetime
provides only the rigid stage for their dynamics. Einstein considered
this lack of back reaction as very unnatural. 

Minkowski spacetime provides the background for relativistic quantum
field theory and the Standard Model of particle physics. In the
non-relativistic limit, it yields quantum mechanics with its absolute,
Newtonian time $t$. This is clearly seen in the Schr\"odinger
equation,
\begin{equation}
\lb{schrodinger}
{\rm i} \hbar \, \frac{\partial \psi}{\partial t} =  \hat{H} \psi\ .
\end{equation}
It must also be noted that the presence of $t$ occurs on the left-hand
side of this equation together with the imaginary unit, i; this fact
will become important below. In relativistic quantum field theory,
\eqref{schrodinger} is replaced by its functional version. 

The Schr\"odinger equation \eqref{schrodinger} is, with respect to
$t$, deterministic and time-reversal invariant. As was already
emphasized by Wolfgang Pauli, the presence of both $t$ and i are
crucial for the probability interpretation of quantum mechanics, in
particular for the conservation of probability {\em in} time.

But the story is not yet complete. It was Einstein's great insight to see
that gravity is a manifestation of the geometry of spacetime; in fact,
gravity is geometry. This led him to his general theory of relativity,
which he completed in 1915. Because of this identification, spacetime
is no longer absolute, but dynamical. There {\em is} now a {\em
  reactio} of all matter and fields onto spacetime and even an
interaction of spacetime with itself (as is e.g. the case in the
dynamics of gravitational waves). 

So, time is absolute in quantum theory, but dynamical in general
relativity. What, then, happens if one seeks a unification of
gravity with quantum theory or, more precisely, seeks an
accommodation of gravity into the quantum framework? Obviously, time
cannot be both absolute and non-absolute: this dilemma is usually referred
to as the ``problem of time'' \cite{Isham,OUP,Kuchar}. 
One can also rephrase it as the problem
of finding a background-independent quantum theory. 

But does one really have to unify gravity with quantum theory into a
theory of quantum gravity? In the next section, I shall give a concise
summary of the main reasons for doing so. I shall then argue that one
can draw important conclusions about the nature of time in quantum gravity 
{\em without} detailed knowledge of the full theory; in fact, all
that is needed is the semiclassical limit -- general relativity. I
shall then describe the approximate nature of any time
parameter and clarify the relevance of these limitations for the
interpretation of quantum theory itself. I shall finally show 
how the direction of time can be understood in a theory which is
fundamentally timeless.


\section{The Disappearance of Time}

The main arguments in favour of quantizing gravity have to do with the
{\em universality} of both quantum theory and gravity. The
universality of quantum theory is encoded in the apparent
universality of the superposition principle, which has passed all
experimental tests so far \cite{deco,Schlosshauer}. There is, of
course, no guarantee that this principle will not eventually break
down. However, I shall make the conservative assumption, in
accordance with all existing experiments, that the superposition
principle does hold universally: arbitrary linear combinations of
physical quantum state do again lead to a physical quantum state; in
general, the resulting quantum states describe highly entangled
quantum systems. If the superposition principle holds universally, it
holds in particular for the gravitational field.   

The universality of the gravitational field is a consequence of its
geometric nature: it couples equally to all forms of energy. It thus
interacts with all quantum states of matter, suggesting 
that it is itself described by a quantum state. This is not a
logical argument, though, but an argument of naturalness \cite{AKR}.  

A further argument for the quantization of gravity is the
incompleteness of general relativity. Under very general assumptions
one can prove singularity theorems that force us to conclude that time
must come to an
end in regions such as the Big Bang and the interior of black
holes. This is, of course, only possible because time in general
relativity is dynamical. The hope, then, is that quantum gravity will
be able to deal with these situations. 

It is generally argued that quantum-gravity effects can only be seen
at a remote scale -- the Planck scale, which originates from the
combination of the three fundamental constants $c$ (speed of light),
$G$ (gravitational constant), and $\hbar$ (quantum of action). The
Planck length, for example, is given by
\be
\lb{lP}
l_{\rm P} = \sqrt{\frac{\hbar G}{c^3}} \approx 1.62 \times 10^{-35} 
                                         \,{\rm m}\ ,
\ee
and is thus much smaller than any length scale that can be probed by the
Large Hadron Collider (LHC). 

This argument is, however, misleading. One may certainly expect that
quantum effects of gravity are always important at the Planck
scale. But they are not restricted to this scale {\em a priori}. The
superposition principle allows the formation of non-trivial
gravitational quantum states at any scale. Why, then, is such a state
not being observed? The situation is analogous to quantum mechanics and
the non-observability of states such as a Schr\"odinger-cat state. And
the reason why such states are not found is the same: decoherence
\cite{deco,Schlosshauer}. The interaction of a quantum system
with its ubiquitous environment (that is, with unaccessible degrees of
freedom) will usually lead to its classical appearance, except for micro-
or mesoscopic situations. The process of decoherence is founded on the
standard quantum formalism, and it has been tested in
many experiments \cite{Schlosshauer}. 

The emergence of classical behaviour through decoherence also holds
for most states of the gravitational field. But there may be
situations where the quantum nature of gravity is visible -- even far
away from the Planck scale. We shall encounter such a situation in
quantum cosmology. It is directly related to the concept of
time in quantum gravity.

Due to the absence of a background structure, the construction of a
quantum theory of gravity is difficult and has not yet been 
accomplished. Approaches are usually divided into two classes. The
more conservative class is the direct quantization of general
relativity; path-integral quantization and canonical quantum gravity
belong to it.
The second class starts from the assumption that a consistent theory
of quantum gravity can only be achieved within a unified quantum
theory of all interactions; superstring theory is the prominent (and
probably unique) example for this class.

In this essay I want to put forward the view that the concept of time
in quantum gravity can be discussed without having the final theory at
one's disposal; the experimentelly tested part of physics together with the
above universality assumptions suffice. 

The arguments are similar in spirit to the ones that led Erwin
Schr\"o\-ding\-er in 1926 to his famous equation \eqref{schrodinger}.  
Motivated by Louis de~Broglie's suggestion of the wave nature of
matter, Schr\"odinger tried to find a wave equation which yields the
equations of classical mechanics in an appropriate limit, in analogy
to the recovery of geometric optics as a limit to the fundamental wave
optics. To achieve this, Schr\"odinger put classical mechanics into
the so-called Hamilton--Jacobi form from which the desired wave equation could
be easily guessed \cite{Schrodinger}.

The same steps can be followed for gravity. One starts by casting
Einstein's field equations into Hamilton--Jacobi form. This was
already done by Asher Peres in 1962 \cite{Peres}. The wave
equation behind the gravitational Hamilton--Jacobi equation is then
nothing but the Wheeler--DeWitt equation, which was derived by John
Wheeler \cite{Wheeler} and Bryce DeWitt \cite{DeWitt} in 1967 from the
canonical formalism. It is of the form
\begin{equation}
\label{wdwfull}
\hat{H}_{\rm tot}\Psi = 0 \ ,
\end{equation}
where $\hat{H}_{\rm tot}$ denotes here the full Hamilton operator for
gravity plus matter. The wave functional $\Psi$ depends on the {\em
  three-dimensional} metric plus all non-gravitational
fields.\footnote{There also exist the so-called diffeomorphism
  constraints, which state that $\Psi$ is independent of the choice of
  spatial coordinates, see e.g. \cite{OUP} for details.} 

The Wheeler--DeWitt equation \eqref{wdwfull} may or may not hold at
the fundamental Planck scale \eqref{lP}. But as long as quantum theory is
universally valid, it will hold at least as an
approximate equation for scales much bigger than $l_{\rm P}$. In this
sense, it is the most reliable equation of quantum gravity, even if it
is not the most fundamental one.

The wave function $\Psi$ in the Wheeler--DeWitt equation
\eqref{wdwfull} does not contain any time parameter $t$. Although at
first glance surprising, this is a straightforward consequence of the
quantum formalism. In classical mechanics, the trajectory of a
particle consists of positions $q$ in time, $q(t)$. In quantum
mechanics, only probability amplitudes for those positions
remain. Because time $t$ is external, the wave function in
\eqref{schrodinger} depends on both $q$ and $t$, but not on any
$q(t)$. In gravity, three-dimensional space is analogous to $q$, and
the classical spacetime corresponds to $q(t)$. Therefore, upon
quantization spacetime vanishes in the same manner as the trajectory
$q(t)$ vanishes. But as there is no absolute time in general
relativity, only space remains, and one is left with \eqref{wdwfull}.  

We can thus draw the conclusion that quantum gravity is 
timeless sole\-ly from the validity of the Einstein equations at large
scales and the assumed universality of quantum theory. Our conclusion
is independent of additional modifications at the Planck scale, such
as the discrete features that are predicted from loop quantum gravity and string
theory.    


\section{Time Regained}

In August 1931, Neville Mott submitted a remarkable paper to the
Cambridge Philosophical Society \cite{Mott}. He discussed the
collision of an alpha-particle with an atom. The remarkable
thing is that he considered the time-independent Schr\"odinger
equation of the total system and used the state of the alpha-particle
to {\em define} time and to derive a time-de\-pen\-dent Schr\"odinger
equation for the atom alone. The total quantum state is of the form
\be
\lb{mott}
\Psi({\bf r},{\bf R})=\psi({\bf r},{\bf R}){\rm e}^{{\rm i}{\bf k}{\bf
    R}}\ ,
\ee
where ${\bf r}$ (${\bf R}$) refers to the atom (alpha-particle). The
time $t$ is then defined from the exponential in \eqref{mott} through
a directional derivative, 
\be
\lb{time}
{\rm i}\frac{\partial}{\partial t}\propto {\rm i}{\bf k}\cdot\nabla_{\bf R}\ .
\ee
This leads to the time-dependent Schr\"odinger equation for the atom.
Such a viewpoint of time as a concept derived from a fundamental
timeless equation is seldom adopted in quantum mechanics. 
It is, however, the key step to understanding the emergence of time from
the timeless Wheeler--DeWitt equation \eqref{wdwfull}. While the
alpha-particle in 
Mott's example corresponds to the gravitational part, the atom
corresponds to the non-gra\-vi\-ta\-tio\-nal degrees of freedom. The time $t$
of the Schr\"odinger equation \eqref{schrodinger} is then {\em
  defined} by a directional derivative similar to
\eqref{time}. Various derivations of such a ``semiclassical time''
have been given in the literature (reviewed e.g. in \cite{OUP}), but
the general idea is always the same. Time emerges from the separation
into two different subsystems: one subsystem (here: the gravitational
part) defines the time with respect to which the other subsystem
(here: the non-gravitational part) evolves.\footnote{More precisely,
  some of the gravitational degrees of freedom can also remain
  quantum, while some of the non-gravitational variables can be
  macroscopic and enter the definition of time.} Time is thus only an
approximate concept. A closer investigation of this approximation
scheme then reveals the presence of quantum-gravitational correction
terms \cite{KS}.

I have remarked above that the Hilbert-space structure of quantum
theory is related to the probability interpretation, and that the
latter seems to be tied to the presence of $t$. In the light of the
fundamental absence of $t$, one may speculate that the Hilbert-space
structure, too, is an approximate structure and that different
mathematical structures are needed for full quantum gravity. 

I have also remarked above that the time $t$ in the Schr\"odinger
equation \eqref{schrodinger} occurs together with the imaginary unit
i. The quantum-mecha\-ni\-cal wave functions are thus complex, which is an
essential feature for the probability interpretation. Since the
Wheeler--DeWitt equation is real, the complex numbers emerge together
with the time $t$ \cite{Barbour2,CK93}. Hasn't this been put in by
hand through the i in the ansatz \eqref{mott}? Not really. One can
start with superpositions of complex wave functions of the form
\eqref{mott}, which together give a real quantum state. But now again
decoherence comes into play. Irrelevant degrees of freedom distinguish
the complex components from each other, making them dynamically
independent \cite{deco}. In a
sense, time is ``measured'' by irrelevant degrees of freedom
(gravitational waves, tiny density fluctuations). 
Some time ago I estimated the magnitude of this effect for a simple 
cosmological model \cite{CK92} and found that the
interference terms between the complex components can be as small as
\be     
\exp\left(-\frac{\pi m
    c^2}{128\hbar H_0}\right)\sim\exp\left(-10^{43}\right)\ ,
\ee
where $H_0$ is the Hubble constant and $m$ the mass of a scalar field,
and some standard numbers have been chosen. This gives further support
for the recovery of time as a viable semiclassical concept. 

There are, of course, situations where the recovery of semiclassical
time breaks down. They can be found through a study of the full
Wheeler--DeWitt equation \eqref{wdwfull}. One can, for example, study
the behaviour of wave packets: semiclassical time is only a viable
approximation if the packets follow the classical trajectory without
significant spreading.
One may certainly expect that a breakdown of the semiclassical limit
occurs at the Planck scale 
\eqref{lP}. But there are 
other situations, too. One occurs for a classically recollapsing
Universe and is described in the next section. Other cases follow from
models with fancy singularities at large scales. The ``big brake'',
for example, corresponds to a Universe which classically comes to an
abrupt halt with infinite deceleration, leading to a singularity at
large scale factor. The corresponding quantum
model was recently discussed in \cite{KKS}. If the wave packet
approaches the classical singularity, the wave function will 
necessarily go to zero there. The time $t$ then loses its meaning, and
all classical evolution comes to an end before the singularity is
reached. One might even speculate that not only time, but also space
disappears \cite{DN}.

The ideas presented here are also relevant to the
interpretation of quantum theory itself. They strongly suggest, for
example, that the Copenhagen interpretation is not applicable in this
domain. The 
reason is the absence of a classical spacetime at the most fundamental
level, which in the Copenhagen interpretation is assumed to exist from the
outset. In quantum gravity, the world is fundamentally timeless and
does not contain classical parts. Classical appearance only emerges
for subsystems through the process of decoherence -- with limitations
dictated by the solution of the full quantum equations.


\section{The Direction of Time}

A fundamental open problem in physics is the origin of irreversibility
in the Universe, the recovery of the arrow of time \cite{Zeh}. 
It is sometimes speculated that this can only be achieved from a
theory of quantum gravity. But can statements about the direction of
time be made if the theory is fundamentally timeless? 

The answer is yes. The clue is, again, the semiclassical nature of the
time parameter $t$. As we have seen in the last section, $t$ is
defined via fundamental gravitational degrees of freedom. The
important point is that the Wheeler--DeWitt equation \eqref{wdwfull}
is {\em asymmetric} with respect to the scale factor that describes
the size of the Universe in a given state. It
assumes a simple form for a small universe, but a complicated form for
a large universe. For small scale factor there is only a minor interaction
between most of the degrees of freedom. The equation then allows the
formulation of a simple 
initial condition \cite{Zeh}: the absence of quantum entanglement
between global degrees of freedom (such as the scale factor) and local
ones (such as gravitational waves or density perturbations). The local
variables serve as an irrelevant environment in the sense of
decoherence.

Absence of entanglement means that the full quantum state is a product
state. Tracing out the environment has then no effect; the state of
the global variables remains pure. There is then no entropy 
(as defined by the reduced density matrix) connected
with them: all information is contained in the system itself.
 The situation changes with increasing scale factor; the
entanglement grows and the entropy for the global variables increases,
too. As soon as the semiclassical approximation is valid, this growth
also holds with respect to $t$; it is inherited from the full
equation. The direction of time is thus defined by the direction of
increasing entanglement. In this sense, the expansion of the universe
would be a tautology. 

There are interesting consequences for a classically recollapsing
universe \cite{KZ}. In order to produce the correct classical limit,
the wave function of the quantum universe must go to zero for large
scale factors. Since the quantum theory cannot distinguish between the
different ends of a classical trajectory (such ends would be the Big Bang and
the Big Crunch), the wave function must consist of many
quasi-classical components with entropies that increase in the
direction of a larger Universe; one could then
never observe a recollapsing universe. In the region where the
classical turning point would be found, all components have to interfere
destructively in order to fulfill the final boundary condition of the
wave function going to zero. This is a drastic example of the
relevance of the superposition principle far away from the Planck
scale -- with possible dramatic consequences for the fate of our
Universe: the classical evolution would come to an end in the future.

Let me emphasize again that 
all the consequences presented in this
essay result from a very conservative starting point: the assumed
universality of quantum theory and its superposition principle. Unless this 
assumption breaks down, these consequences should hold in every
consistent quantum theory of gravity. 
We are able to understand from the fundamental picture of a timeless
world both the emergence and the limit of our usual concept of time. 

\vskip 3mm

I thank Marcel Reginatto and H.-Dieter Zeh for their comments on this
manuscript. 




\begin{thebibliography}{99}

\bibitem{Einstein} A. Einstein, How I created the theory of
       relativity. Translated by Y. A. Ono. {\em Physics Today},
       August~1982, 45--47. 

\bibitem{Barbour} J. B. Barbour, Leibnizian time, Machian dynamics,
           and quantum gravity. In: {\em Quantum concepts in space and
           time}, edited by R. Penrose and C. J. Isham
           (Oxford University Press, Oxford, 1986), pp.~236--246. 

\bibitem{Isham} C. J. Isham, Canonical quantum gravity and the problem of
           time. In: {\em Integrable systems, quantum groups, and
           quantum field theory}, edited by L. A. Ibort and
         M. A. Rodr\'{\i}guez) (Kluwer, Dordrecht, 1993),
           pp.~157--287. 

\bibitem{OUP} C. Kiefer, {\em Quantum Gravity}, second edition (Oxford
              University Press, Oxford, 2007).

\bibitem{Kuchar} K. V. Kucha\v{r}, Time and interpretations of quantum
           gravity. In: {\em Proceedings of the 4th Canadian
             Conference on 
           General Relativity and Relativistic Astrophysics}, edited by
           G. Kunstatter, D. Vincent, and J. Williams
           (World Scientific, Singapore, 1993),
            pp.~211--314.

\bibitem{deco} E. Joos, H. D. Zeh, C. Kiefer, D. Giulini,
J. Kupsch, and I.-O. Stamatescu, {\em Decoherence and the
Appearance of a Classical World in Quantum Theory}, second edition
(Springer, Berlin, 2003).

\bibitem{Schlosshauer} M. Schlosshauer, {\em Decoherence and the
    Quantum-to-Classical Transition} (Springer, Berlin, 2007).

\bibitem{AKR} M. Albers, C. Kiefer, and M. Reginatto,
Measurement analysis and quantum gravity. {\em Physical Review D},
{\bf 78}, 064051 (2008). 

\bibitem{Schrodinger} E. Schr\"odinger, Quantisierung als
  Eigenwertproblem II.
                       {\em Annalen der Physik} {\bf 384}, 489--527 (1926). 

\bibitem{Peres} A. Peres, On Cauchy's problem in general relativity--II.
           {\em Nuovo Cimento}, {\bf XXVI}, 53--62 (1962).

\bibitem{Wheeler} J. A. Wheeler, Superspace and the nature of quantum
           geometrodynamics. In: {\em Battelle rencontres}, edited by
           C. M. DeWitt and J. A. Wheeler (Benjamin, New York, 1968),
          pp.~242--307. 

\bibitem{DeWitt} B. S. DeWitt, Quantum theory of gravity. I.
           The canonical theory. {\em Physical Review}, {\bf 160},
           1113--1148 (1967).

\bibitem{Mott} N. F. Mott, On the theory of excitation by collision with
           heavy particles. {\em Proceedings of the Cambridge
             Philosophical Society}, {\bf 27}, 
           553--560 (1931).

\bibitem{KS} C. Kiefer and T. P. Singh, Quantum gravitational
           correction terms to the functional Schr\"odinger equation.
           {\em Physical Review D}, {\bf 44}, 1067--1076 (1991).

\bibitem{Barbour2} J. B. Barbour, Time and complex numbers in canonical
           quantum gravity. {\em Physical Review D}, {\bf 47},
           5422--5429 (1993).  

\bibitem{CK93} C. Kiefer, Topology, decoherence, and semiclassical
  gravity. {\em Physical Review D}, {\bf 47}, 5414--5421 (1993).

\bibitem{CK92} C. Kiefer, Decoherence in quantum electrodynamics
           and quantum gravity. {\em Physical Review D}, {\bf 46},
           1658--70 (1992).

\bibitem{KKS} A. Y. Kamenshchik, C. Kiefer, and B. Sandh\"ofer,
  Quantum cosmology with a big-brake singularity.
 {\em Physical Review D}, {\bf 76}, 064032 (2007).

\bibitem{DN} T. Damour and H. Nicolai, Symmetries, singularities and
  the de-emergence of space. {\em International Journal of Modern
    Physics D}, {\bf 17}, 525--531 (2008).

\bibitem{Zeh} H. D. Zeh, {\em The physical basis of the direction
           of time}, fifth edition (Springer, Berlin, 2007).

\bibitem{KZ} C. Kiefer and H. D. Zeh, Arrow of time in a recollapsing
           quantum universe. {\em Physical Review D}, {\bf 51},
           4145--4153 (1995). 


\end{thebibliography}
\end{document}